\documentclass[11pt]{article}
\usepackage{amsmath,amssymb,amsfonts,fancyhdr,fancybox,graphics,epsfig,calc,color,subfigure}
\usepackage[pagewise, right]{lineno}

\renewcommand{\baselinestretch}{1.4}
\def\blist#1#2#3
{    \begin{list}{}{
     \setlength{\parsep}{0pt}
     \setlength{\leftmargin}{#1 pt}
     \setlength{\listparindent}{0pt}
     \setlength{\itemindent}{\listparindent}
     \setlength{\labelsep}{#3 pt}
     \setlength{\labelwidth}{\leftmargin}
     \addtolength{\labelwidth}{-\labelsep}
     \addtolength{\labelwidth}{\itemindent}
     \setlength{\rightmargin}{#2 pt}   }   }

\def\elist{\end{list}}

\newtheorem{proposition}{Proposition}

\renewcommand{\baselinestretch}{1.38}
\usepackage{amsmath}
\def\boxit#1{\vbox{\hrule\hbox{\vrule\kern6pt\vbox{\kern6pt#1\kern6pt}\kern6pt\vrule}\hrule}}

\begin{document}

\topmargin -0.3in \oddsidemargin 0.2in \evensidemargin 0.2in
\baselineskip 9mm
\renewcommand \baselinestretch {1.2}

\title{\bf \large A note on selection stability: combining stability and prediction }

\date{}

\author{\begin{tabular}{c} Yixin Fang\footnote{Correspondence to: 650 First Avenue, Room 559, New York, NY 10016; yixin.fang@nyumc.org}\\
\emph{New York University School of Medicine} \\\\
Junhui Wang\\\emph{University of Illinois at Chicago}
\\\\
Wei Sun\\\emph{Purdue University}
\end{tabular}}

\titlepage

\maketitle

\begin{center}
\begin{minipage}{132mm}
\begin{center}{\bf Abstract}\end{center}
Recently, many regularized procedures have been proposed for variable selection in linear regression, but their performance depends on the tuning parameter selection. Here a criterion for the tuning parameter selection is proposed, which combines the strength of both stability selection and cross-validation and therefore is referred as the prediction and stability selection (PASS). The selection consistency is established assuming the data generating model is a subset of the full model, and the small sample performance is demonstrated through some simulation studies where the assumption is either held or violated.

\end{minipage}
\end{center}

\bigskip


\newpage
\setcounter{equation}{0}

\linenumbers

\section{Introduction}

Many regularized procedures produce sparse solution and therefore are sometimes used for variable selection in linear regression. Breiman (1996) showed that regularized procedures are more stable than subset selection. Such procedures include LASSO (Tibshirani, 1996), SCAD (Fan and Li, 2001), and adaptive LASSO (Zou, 2006). However, their performance depends crucially on the tuning parameter selection.

This manuscript is not intended to add a new regularized procedure to the long list. Rather, it is aimed to propose a new method for selecting an ``appropriate" tuning parameter, which is crucial in any existing regularized procedure. The meaning of appropriateness depends on whether the purpose of regularization is prediction or variable selection.

For prediction, popular methods for the tuning parameter selection include $C_p$ (Mallows, 1973), cross-validation (Stone, 1974), and generalized cross-validation (Craven and Wahba, 1979). However, for prediction, it is way too simple to consider only one regularization procedure based on one selected tuning parameter, and usually it is more powerful to consider complicated procedures such as boosting and averaging (Hastie {\it et al.},~2009). Therefore, this manuscript is focused on the tuning parameter selection for variable selection.

For variable selection, the most popular method for the tuning parameter selection is BIC (Schwarz, 1978). The selection consistency of BIC for SCAD was shown in several papers (e.g., Wang {\it et al.},~2007, Wang {\it et al.},~2009, and Zhang {\it et al.},~2010). Here the selection consistency means that the probability of selecting the data generating model is tending to one when the sample size goes to infinity, assuming that the data generating model is a subset of the full model. This manuscript is to propose an alternative to BIC. The new method is selection consistent for a large group of regularized procedures.

Simple put, the new method combines the strength of both stability selection and cross-validation, and therefore it is referred as the prediction and stability selection (PASS). Here the stability selection is a recent idea for variable selection. Bach (2008) proposed Bolasso to enhance the original LASSO through the bootstrap; but it requires knowing the exact root-$n$ regularization decay. Meinshausen and B$\ddot{\mbox{u}}$hlmann (2010) proposed their version of stability selection, in which a super tuning parameter, cutoff $\pi_{thr}$ (pre-set as 0.8 there), needs to be selected. Most recently, Sun {\it et al.}~(2012) proposed Kappa selection; however, there is also a super tuning parameter, threshold $\alpha_n$ (pre-set as 0.1 there), needed to be selected.

This manuscript is a note on Sun {\it et al.}~(2012), aimed at avoiding the selection of threshold $\alpha_n$ by incorporating the strength of cross-validation. The remainder of the manuscript is organized as follows. Section 2 reviews some asymptotic results in some regularized procedures. Section 3 develops a new criterion for tuning parameter selection and Section 4 examines its selection consistency. Numerical results are in Section 5 and some discussion is in Section 6.

\section{Regularized procedures}

Consider variable selection in linear regression,
\begin{equation}
y_i=x_i'\beta+\epsilon_i,\ \ i=1, \cdots, n, \label{linear_model}
\end{equation}
where $\beta=(\beta_1, \cdots, \beta_p)'$, $E(\epsilon_i)=0$, and $Var(\epsilon_i)=\sigma^2$. Assume both response and covariates are centered and then no intercept is included. Let $\mathcal{A}=\{j:
\beta_j\neq 0\}$ and assume $\beta$ is sparse in the sense that $|\mathcal{A}|=q<p$. Without loss of generality, assume $\mathcal{A}=\{1, \cdots, q\}$.

A general framework for the regularized regression is
\begin{equation}
\widehat{\beta}_{\lambda}=\arg\min_{\gamma\in {\mathbb R}^p}\sum_{i=1}^n(y_i-x_i'\gamma)^2/n+\sum_{j=1}^pp_{\lambda}(|\gamma_j|), \label{regularization}
\end{equation}
where $p_{\lambda}(\cdot)$ is a regularization term encouraging sparsity in $\widehat{\beta}$. In LASSO, $p_{\lambda}(|\beta_j|)=\lambda|\beta_j|$. In SCAD, $p'_{\lambda}(\theta)=\lambda\{I(\theta\leq\lambda)+\frac{(a\lambda-\theta)_{+}}{(a-1)\lambda}I(\theta>\lambda)\}$. And in adaptive LASSO, $p_{\lambda}(|\beta_j|)=\lambda|\beta_j|/|\widetilde{\beta}_j|$, where $\widetilde{\beta}_j$ is some initial estimate of $\beta_j$.

If $\widehat{\mathcal{A}}_{\lambda}=\{j: \widehat{\beta}_{\lambda j}\neq 0\}$ is used to estimate $\mathcal{A}$, all the three aforementioned regularization procedures have been shown to be selection consistent under various conditions with appropriately $\lambda=\lambda_n$, where subscript $n$ emphasize the dependence on sample size $n$.


For simplification, in this manuscript, consider the case where $p$ is fixed. It has been shown that for all these three regularization procedures, there exist $r_n$ and $s_n$ such that the procedures are selection consistent if $r_n\prec \lambda_n \prec s_n$, where $a_n\prec b_n$ means $a_n=o(b_n)$. This fact might also hold for many other regularization procedures. Specifically, for LASSO under the irrepresentable condition, $r_n\asymp 1/\sqrt{n}$ and $s_n\asymp 1$ (Zhao and Yu, 2006), where $a_n\asymp b_n$ means $a_n=O(b_n)$ and $b_n=O(a_n)$. In addition, $r_n\asymp 1/\sqrt{n}$ and $s_n\asymp 1$ for SCAD (Fan and Li, 2001) and $r_n\asymp 1/n$ and $s_n\asymp 1/\sqrt{n}$ for adaptive LASSO (Zou, 2006). In the following, five mutually exclusive cases of $\lambda_n$ are considered. For LASSO, refer to Bach (2008), while for the other two, refer to Sun {\it et al.}~(2012).

{\it Case 1:} If $\lambda_n\succ s_n$, then $\widehat{\beta}_{\lambda_n}={\bf 0}$ with probability tending to one.

{\it Case 2:} If $\lambda_n\asymp s_n$, then $\widehat{\beta}_{\lambda_n}\rightarrow {\gamma_0}\neq \beta$, where $\gamma_0$ is fixed and its sign pattern may or may not be the same as that of $\beta$.

{\it Case 3:} If $r_n\prec \lambda_n \prec s_n$, then $\widehat{\beta}_{\lambda_n}\rightarrow \beta$ and the sign pattern of $\widehat{\beta}_{\lambda_n}$ is consistent with that of $\beta$ with probability tending to one. Here the irrepresentable condition is needed for LASSO but not for the other two.

{\it Case 4:} If $\lambda_n\asymp r_n$, then the sign pattern of $\widehat{\beta}_{\lambda_n}$ is consistent with that of $\beta$ on  $\mathcal{A}$ with probability tending to one, while for all sign patterns consistent with that of $\beta$ on $\mathcal{A}$, the probability of obtaining this pattern is tending to a limit in $(0, 1)$.

{\it Case 5:} If $\lambda_n\prec r_n$, then $\widehat{\beta}_{\lambda_n}\rightarrow {\beta}$ and $\widehat{\mathcal{A}}_{\lambda_n}=\{1, \cdots, p\}$ with probability tending to one.

\section{Prediction and stability selection (PASS)}

A good criterion should intend to select $\lambda_n$ from case 3; selecting $\lambda_n$ from cases 1 or 2 might lead to under-fitting while from cases 4 or 5 might lead to over-fitting. If the two degenerate cases (1 and 5) are pre-excluded, the criterion designed in this section incorporates cross-validation, which avoids under-fitting, and Kappa selection proposed in Sun {\it et al.}~(2012), which avoids over-fitting.

To describe this criterion, consider any aforementioned regularized procedure with tuning parameter $\lambda$. First of all, randomly partition the dataset $\{(y_i, x_i), \cdots, (y_n, x_n) \}$ into two halves, $Z_1=\{(y^{*}_1, x^{*}_1), \cdots, (y^{*}_m, x^{*}_m)\}$ and $Z_2=\{(y^{*}_{m+1}, x^{*}_{m+1}), \cdots, (y^{*}_n, x^{*}_n)\}$, where $m=\lfloor n/2 \rfloor$.  Based on $Z_1$ and $Z_2$ respectively, $\widehat{\beta}_{k\lambda}$ is obtained via (\ref{regularization}) and then submodel $\widehat{\mathcal{A}}_{k\lambda}$ is selected, $k=1, 2$.

If $\lambda$ were from Case 4, both submodels, $\widehat{\mathcal{A}}_{k\lambda}, k=1, 2,$ would include non-informative variables randomly. The agreement of these two submodels can be measured by Cohen's Kappa Coefficient (Cohen, 1960),
\begin{equation}
\kappa(\widehat{\mathcal{A}}_{1\lambda}, \widehat{\mathcal{A}}_{2\lambda})=\frac{Pr(a)-Pr(e)}{1-Pr(e)},\label{Kappa}
\end{equation}
where $Pr(a)=(|\widehat{\mathcal{A}}_{1\lambda} \cap \widehat{\mathcal{A}}_{2\lambda}|+|\widehat{\mathcal{A}}_{1\lambda}^{c} \cap \widehat{\mathcal{A}}_{2\lambda}^{c}|)/p$ and $Pr(e)=(|\widehat{\mathcal{A}}_{1\lambda}||\widehat{\mathcal{A}}_{2\lambda}|+
|\widehat{\mathcal{A}}_{1\lambda}^{c}||\widehat{\mathcal{A}}_{2\lambda}^{c}|)/p^2$.

On the other hand, if $\lambda$ were from Case 2, either submodels, $\widehat{\mathcal{A}}_{k\lambda}, k=1, 2,$ might exclude some informative variable. To avoid such under-fitting, consider cross-validation,
\begin{equation}
CV(Z_1, Z_2; \lambda)=\{\sum_{i=1}^m(y_i-x_i'\widehat{\beta}_{2\lambda})^2+
\sum_{i=m+1}^n(y_i-x_i'\widehat{\beta}_{1\lambda})^2\}/n. \label{CV}
\end{equation}

In addition, submodel $\mathcal{A}$ is assumed to be sparse and contain at least one variable, so $\kappa(\widehat{\mathcal{A}}_{1\lambda}, \widehat{\mathcal{A}}_{2\lambda})$ will be set as $-1$ if both $\widehat{\mathcal{A}}_{1\lambda}$ and $\widehat{\mathcal{A}}_{2\lambda}$ are empty or both are full (that is, the two degenerate cases, Cases 1 and 5, will be pre-excluded).

%

Now we are ready to describe the PASS algorithm, which runs the following five steps.

{\it Step 1:} Randomly partition the original dataset into two halves, $Z_1^{*b}$ and $Z_2^{*b}$.

{\it Step 2:} Based on $Z_1^{*b}$ and $Z_2^{*b}$ respectively, two sub-models, $\widehat{\mathcal{A}}^{*b}_{1\lambda}$ and $\widehat{\mathcal{A}}^{*b}_{1\lambda}$, are selected.

{\it Step 3:} Calculate $\kappa(\widehat{\mathcal{A}}^{*b}_{1\lambda}, \widehat{\mathcal{A}}^{*b}_{2\lambda})$ and $CV(Z^{*b}_1, Z^{*b}_2; \lambda)$.

{\it Step 4:} Repeat Steps 1-3 for $B$ times and obtain the following ratio,
\begin{equation}
PASS(\lambda)=\sum_{b=1}^B \kappa(\widehat{\mathcal{A}}^{*b}_{1\lambda}, \widehat{\mathcal{A}}^{*b}_{2\lambda})/\sum_{b=1}^B CV(Z^{*b}_1, Z^{*b}_2; \lambda). \label{PASS}
\end{equation}

{\it Step 5:} Compute $PASS(\lambda)$ on a grid of $\lambda$ and select $\widehat{\lambda}=\arg\max_{\lambda}PASS(\lambda)$.

\section{Selection consistency}

Recall the existence of those $r_n$ and $s_n$ in Section 2, which plays an important role here. The underlying assumptions are not stated in the following theorem, but they can be found in Bach (2008) for LASSO, Fan and Li (2001) for SCAD, and Zou (2006) for adaptive LASSO. As discussed in Section 3, Cases 1 and Case 5 can be pre-excluded by the definition of $\kappa$, so it suffices to show that the PASS can distinguish Case 3 from Cases 2 and 4.

{\proposition  For any $\lambda_n$ such that $r_n\prec\lambda_n\prec s_n$, as $n\rightarrow\infty$ and $B\rightarrow\infty$,
$$Pr\{PASS(s_n)<PASS(\lambda_n)\}\rightarrow 1 {\mbox \ and\ } Pr\{PASS(r_n)<PASS(\lambda_n)\}\rightarrow 1. $$}
\noindent {\it Heuristic proof:} First, by Chebyshev's inequality, for identically distributed variables, $X_{nb}, b=1, \cdots, B$, if $Var(X_{nb})<C$ and $Corr(X_{n1}, X_{n2})\rightarrow 0$, then $\sum_{b=1}^BX_{nb}/B-E(X_{n1})\rightarrow 0$.

If $r_n\prec\lambda_n\prec s_n$, by the result in Case 3, $\widehat{\mathcal{A}}^{*1}_{1\lambda_n}=\widehat{\mathcal{A}}^{*1}_{2\lambda_n}=\mathcal{A}$ with probability tending to one, and therefore by Lebesgue's dominated theorem, $E\{\kappa(\widehat{\mathcal{A}}^{*1}_{1\lambda_n}, \widehat{\mathcal{A}}^{*1}_{2\lambda_n})\}\rightarrow 1$. In order to apply Lebesgue's dominated theorem to examine the asymptotic property of cross-validation, assume that in (\ref{CV}), $\widehat{\beta}_{2\lambda}$ and $\widehat{\beta}_{2\lambda}$ are bounded manually by some large value, say $M=10^6$. Then, by Lebesgue's dominated theorem, $E\{CV(Z^{*1}_1, Z^{*1}_2; \lambda_n )\} \rightarrow \sigma^2$.

When $\lambda=s_n$, $E\{CV(Z^{*1}_1, Z^{*1}_2; s_n )\} \rightarrow \sigma^2+c||\beta-\gamma_0||^2$, where $\gamma_0$ is defined in Case 2 and $c$ is the limit of the minimum eigenvalue of $\sum x_ix_i'/n$ (fixed design matrix) or the minimum eigenvalue of $E(x_1x_1')$ (random design matrix). And trivially, $\kappa(\widehat{\mathcal{A}}^{*1}_{1s_n}, \widehat{\mathcal{A}}^{*1}_{1s_n})\leq 1$. Therefore, $Pr\{PASS(s_n)<PASS(\lambda_n)\}\rightarrow 1$.

When $\lambda=r_n$, by the result in Case 4, $Pr(\widehat{\mathcal{A}}^{*1}_{1r_n}\neq \widehat{\mathcal{A}}^{*1}_{2r_n})\rightarrow \delta >0$. Note that if $\widehat{\mathcal{A}}^{*1}_{1r_n}\neq \widehat{\mathcal{A}}^{*1}_{2r_n}$, then $\kappa(\widehat{\mathcal{A}}^{*1}_{1r_n}, \widehat{\mathcal{A}}^{*1}_{2r_n})\leq 1-1/p$. Then $\lim_{n\rightarrow\infty} E\{\kappa(\widehat{\mathcal{A}}^{*1}_{1\lambda_n}, \widehat{\mathcal{A}}^{*1}_{2\lambda_n})\}\leq (1-\delta)+(1-1/p)\delta<1$. And trivially, $E\{CV(Z^{*1}_1, Z^{*1}_2; r_n )\} \rightarrow \sigma^2$. Therefore, $Pr\{PASS(r_n)<PASS(\lambda_n)\}\rightarrow 1$. $\Box$

\section{Numerical results}

In this section, via simulations, the PASS method is compared with Cp, 10-fold cross-validation (CV), generalized cross-validation (GCV), and BIC. AIC is not compared because it is equivalent to Cp here. R package pass is created for implementing both the PASS method proposed here and the Kappa selection method proposed in Sun et al. (2012). After $\widehat{\lambda}$ is selected by one of the above criterions, submodel $\widehat{\mathcal{A}}_{\widehat{\lambda}}$ is selected based on the non-zero components of $\widehat{\beta}_{\widehat{\lambda}}$ obtained from (\ref{regularization}). In addition, the OLS estimate based on only the selected variables, $\widetilde{\beta}$, is also obtained, along with its relative prediction error, $RPE=E(x_0'\widetilde{\beta}-x_0'\beta)^2/\sigma^2$, where $x_0$ is i.i.d. with $x_i$.

Three scenarios are considered. In Scenario I, the data generating model is a subset of the full model. In Scenario II, tapering effects are added to the generating model in Scenario I. In Scenario III, the dimension of the data increases with the sample size.

In Scenario I, the data generating model is (\ref{linear_model}) where $\beta=(3, 1.5, 0, 0, 2, 0, 0, 0)'$, and $x_{i1}, \cdots, x_{ip}$ and $\epsilon_i$ are generated from $N(0,1)$ with $Corr(x_{ik}, x_{il})=0.5^{|k-l|}$. This example was commonly used in literature, such as Tibshirani (1996), Fan and Li (2001), and Zou (2006). Sample size $n$ is set as $40$, $60$, and $80$. Three regularized procedures are applied, LASSO, adaptive LASSO (aLASSO), and SCAD. Tuning parameter $\lambda$ are searched among $\{10^{-2+4k/99}; k=0, \cdots, 99\}$. The number of random partitions is set as $B=20$.

Each simulation setting is repeated $100$ times. The percentage of selecting the sparse generating model $\mathcal{A}=\{1, 2, 5\}$ and the relative prediction error (RPE) of the selected submodel are summarized in Table 1. The average numbers of correctly selected zeros (C) and incorrectly selected zeros (I) are summarized in Table 2.

\begin{table}[htp]
\caption{Percentage (PCT) of selecting $\{1, 2, 5\}$ and average RPE of selected submodels}
\label{tab1}
\vskip 10pt
\centering
\begin{tabular}{cr|cccccccccc}
\hline
 & & \multicolumn{2}{c}{PASS} & \multicolumn{2}{c}{BIC} &\multicolumn{2}{c}{$C_p$}& \multicolumn{2}{c}{CV} &\multicolumn{2}{c}{GCV} \\
 \hline
 $n$  & Method & PCT & RPE  & PCT & RPE  & PCT & RPE  & PCT &RPE  & PCT & RPE  \\
 \hline
   & LASSO  & 0.45 & 0.142 & 0.29 & 0.183 & 0.16 & 0.203 & 0.09 & 0.220 & 0.16 & 0.203\\
40 & aLASSO & 0.94 & 0.102 & 0.75 & 0.143 & 0.53 & 0.181 & 0.63 & 0.167 & 0.52 & 0.181\\
   & SCAD   & 0.99 & 0.092 & 0.81 & 0.141 & 0.55 & 0.180 & 0.76 & 0.152 & 0.52 & 0.184\\
\hline
   & LASSO  & 0.49 & 0.095 & 0.35 & 0.112 & 0.16 & 0.138 & 0.14 & 0.140 & 0.17 & 0.137\\
60 & aLASSO & 0.99 & 0.069 & 0.87 & 0.084 & 0.52 & 0.118 & 0.65 & 0.103 & 0.52 & 0.118\\
   & SCAD   & 1.00 & 0.066 & 0.88 & 0.084 & 0.58 & 0.118 & 0.76 & 0.100 & 0.56 & 0.119\\
\hline
   & LASSO  & 0.60 & 0.055 & 0.38 & 0.074 & 0.16 & 0.097 & 0.08 & 0.097 & 0.16 & 0.098\\
80 & aLASSO & 0.99 & 0.042 & 0.88 & 0.056 & 0.56 & 0.081 & 0.77 & 0.067 & 0.56 & 0.081\\
   & SCAD   & 0.99 & 0.044 & 0.89 & 0.056 & 0.62 & 0.079 & 0.75 & 0.069 & 0.61 & 0.080\\
\hline

\end{tabular}
\end{table}

\begin{table}[htp]
\caption{Average numbers of correctly selected zeros (C) and incorrectly selected zeros (I)}
\label{tab2}
\vskip 10pt
\centering
\begin{tabular}{cr|cccccccccc}
\hline
 & & \multicolumn{2}{c}{PASS} & \multicolumn{2}{c}{BIC} &\multicolumn{2}{c}{$C_p$}& \multicolumn{2}{c}{CV} &\multicolumn{2}{c}{GCV} \\
 \hline
 $n$  & Method & C & I  & C & I  & C & I  & C & I  & C & I  \\
 \hline
   & LASSO  & 4.16 & 0 & 3.68 & 0 & 3.26 & 0 & 2.66 & 0 & 3.25 & 0 \\
40 & aLASSO & 4.94 & 0 & 4.59 & 0 & 4.16 & 0 & 4.25 & 0 & 4.15 & 0 \\
   & SCAD   & 4.99 & 0 & 4.63 & 0 & 4.11 & 0 & 4.39 & 0 & 4.06 & 0 \\
\hline
   & LASSO  & 4.36 & 0 & 4.00 & 0 & 3.12 & 0 & 2.85 & 0 & 3.13 & 0 \\
60 & aLASSO & 4.99 & 0 & 4.84 & 0 & 4.17 & 0 & 4.35 & 0 & 4.17 & 0 \\
   & SCAD   & 5.00 & 0 & 4.84 & 0 & 4.15 & 0 & 4.37 & 0 & 4.12 & 0 \\
\hline
   & LASSO  & 4.47 & 0 & 4.05 & 0 & 3.01 & 0 & 2.66 & 0 & 3.00 & 0 \\
80 & aLASSO & 4.99 & 0 & 4.84 & 0 & 4.19 & 0 & 4.49 & 0 & 4.19 & 0 \\
   & SCAD   & 4.99 & 0 & 4.83 & 0 & 4.23 & 0 & 4.45 & 0 & 4.22 & 0 \\
\hline

\end{tabular}
\end{table}

Table 1 shows that PASS performs much better than the other criterions in terms of having the largest percentage of selecting submodel $\mathcal{A}=\{1, 2, 5\}$. In addition, if the selected model is used for prediction (although in practice it is way too simple for prediction), PASS performs better than the others in terms of having the smallest RPE. It also verifies that, in terms of variable selection, adaptive LASSO and SCAD perform better than LASSO and BIC performs better than $C_p$, CV, and GCV. Furthermore, Table 2 shows that all the criterions barely (never happen in the 100 times here) select any incorrect zeros. It seems PASS performs much better than the others in terms of selecting the largest number of correct zeros; there are 5 correct zeros in the data generating model.

In Scenario II, the consequence of adding tapering effects is examined. Three generating models are considered: (II.1) $\beta=(3, 2, 1.5, 0.05, 0.04, 0.03, 0.02, 0.01)'$; (II.2) $\beta=(3, 2, 1.5, 0.1, 0.08, 0.06, 0.04, 0.02)'$; and (II.3) $\beta=(3, 2, 1.5, 0.2, 0.16, 0.12, 0.08, 0.04)'$. Other setups are the same as those in Scenario I except that sample size $n$ is set as $40$. Table 3 summarizes the average size and the average RPE of the selected submodels.

\begin{table}[htp]
\caption{Average size and average RPE of selected submodels}
\label{tab3}
\vskip 10pt
\centering
\begin{tabular}{cr|cccccccccc}
\hline
 & & \multicolumn{2}{c}{PASS} & \multicolumn{2}{c}{BIC} &\multicolumn{2}{c}{$C_p$}& \multicolumn{2}{c}{CV} &\multicolumn{2}{c}{GCV} \\
 \hline
 Model  & Method & Size & RPE  & Size & RPE  & Size & RPE  & Size &RPE  & Size & RPE  \\
 \hline
   & LASSO    & 3.58 & 0.145 & 4.08 & 0.179 & 4.83 & 0.208 & 5.15 & 0.212 & 4.79 & 0.207\\
II.1 & aLASSO & 3.08 & 0.133 & 3.33 & 0.149 & 3.88 & 0.187 & 3.83 & 0.174 & 3.91 & 0.189\\
   & SCAD     & 3.08 & 0.122 & 3.27 & 0.143 & 3.89 & 0.191 & 3.71 & 0.169 & 3.98 & 0.197\\
\hline
   & LASSO    & 3.80 & 0.166 & 4.58 & 0.194 & 5.06 & 0.208 & 5.52 & 0.211 & 5.06 & 0.209\\
II.2 & aLASSO & 3.17 & 0.172 & 3.63 & 0.184 & 4.22 & 0.208 & 4.10 & 0.200 & 4.21 & 0.208\\
   & SCAD     & 3.20 & 0.158 & 3.51 & 0.178 & 4.16 & 0.206 & 3.89 & 0.191 & 4.19 & 0.209\\
\hline
   & LASSO    & 4.54 & 0.215 & 5.28 & 0.211 & 5.75 & 0.222 & 6.25 & 0.224 & 5.69 & 0.220\\
II.3 & aLASSO & 3.36 & 0.260 & 4.17 & 0.249 & 4.71 & 0.230 & 4.67 & 0.246 & 4.72 & 0.230\\
   & SCAD     & 3.59 & 0.252 & 4.16 & 0.253 & 4.74 & 0.234 & 4.69 & 0.249 & 4.75 & 0.235\\
\hline

\end{tabular}
\end{table}

Table 3 shows that PASS is more immune to tapering effects than the other criterions. In model (II.1), the signal-to-noise ratio (SNR) of the largest tapering effect is $0.05/\sqrt{40}=0.316$, and therefore it is desirable to exclude all 5 tapering effects and PASS outperforms the other. In model (II.2), the SNP of the largest tapering effect is $0.1/\sqrt{40}=0.632$, and still it is reasonable to exclude all 5 tapering effects and PASS outperforms the other. However, in model (II.3), the SNP of the largest tapering effect is $0.2/\sqrt{40}=1.264$, and therefore it is arguable to exclude all 5 tapering effects. Still, PASS selects sparser submodels than the others, but in some cases PASS has slightly bigger RPE than others.

In Scenario III, we investigate the effects of the dimensionality. The setting is similar to the one in Scenario I except that $\beta=(5,4,3,2,1,0,\cdots,0)^T$, $p=[\sqrt{n}]$. More specifically, $3$ cases are examined: (1) $n=100,~ p=10$; (2) $n=200, ~p=14$; and (3) $n=400, ~p=20$. The percentage of selecting the sparse generating model $\mathcal{A}=\{1, 2, 3, 4, 5\}$ and the relative prediction error (RPE) of the selected submodel are summarized in Table $\ref{tab4}$.

\begin{table}[htp]
\caption{Percentage (PCT) of selecting $\{1, 2, 3, 4, 5\}$ and average RPE of selected submodels}
\label{tab4}
\vskip 10pt
\centering
\begin{tabular}{cr|cccccccccc}
\hline
 & & \multicolumn{2}{c}{PASS} & \multicolumn{2}{c}{BIC} &\multicolumn{2}{c}{$C_p$}& \multicolumn{2}{c}{CV} &\multicolumn{2}{c}{GCV} \\
 \hline
 $n(p)$& Method & PCT & RPE  & PCT & RPE  & PCT & RPE  & PCT &RPE  & PCT & RPE  \\
 \hline
        & LASSO  & 0.74 & 0.055 & 0.43 & 0.083 & 0.17 & 0.084 & 0.10 & 0.082 & 0.17 & 0.084\\
100(10) & aLASSO & 0.96 & 0.049 & 0.86 & 0.053 & 0.48 & 0.061 & 0.74 & 0.056 & 0.47 & 0.063\\
        & SCAD   & 0.97 & 0.048 & 0.92 & 0.049 & 0.47 & 0.073 & 0.82 & 0.050 & 0.47 & 0.072\\
\hline
        & LASSO  & 0.89 & 0.022 & 0.49 & 0.040 & 0.11 & 0.043 & 0.07 & 0.045 & 0.11 & 0.043\\
200(14) & aLASSO & 0.99 & 0.018 & 0.90 & 0.024 & 0.38 & 0.037 & 0.66 & 0.027 & 0.38 & 0.037\\
        & SCAD   & 1.00 & 0.018 & 0.93 & 0.022 & 0.46 & 0.038 & 0.73 & 0.024 & 0.47 & 0.038\\
\hline
        & LASSO  & 0.95 & 0.012 & 0.53 & 0.029 & 0.09 & 0.025 & 0.04 & 0.023 & 0.09 & 0.025\\
400(20) & aLASSO & 1.00 & 0.012 & 0.93 & 0.013 & 0.34 & 0.019 & 0.73 & 0.015 & 0.33 & 0.019\\
        & SCAD   & 1.00 & 0.012 & 0.98 & 0.012 & 0.43 & 0.020 & 0.75 & 0.012 & 0.43 & 0.021\\
\hline
\end{tabular}
\end{table}

%
%

Clearly the proposed PASS criterion outperforms other competitors in both variable selection and prediction performance. As illustrated in Table \ref{tab4}, PASS delivers the largest percentage of selecting the true active set among all the selection criteria, and yields the smallest relative prediction error across all cases.

\section{Discussion}

In literature, BIC is commonly used for tuning parameter selection in regularized procedures. Recently, stability selection is becoming popular. The intuition behind stability selection is that a good variable selection criterion should select similar subsets of variables when applied to different samples of data generated from a same population. However, if there were a few variables of significantly large effects, then any selection criterion selecting only these ``big" variables would be stable, and therefore applying stability selection would lead to under-fitting. The PASS criterion proposed here overcomes this drawback by borrowing the strength from cross-validation.

Although it is showed that the PASS criterion is selection consistent, it is worth noting that selection consistency is meaningful only in theory because a naively simple true model is assumed. In practice, it is extremely important to evaluate carefully scientific aspects of the full model before conducting variable selection. Practically, the PASS, along with many other criteria, can be only treated as tools for data mining or data dredging. In other words, these variable selection criteria are exploratory rather than confirmatory.

Another limitation of the proposed criterion, although it is only technical, is that the selected $\widehat{\lambda}$ is corresponding to sample size $n/2$, because each time data are partitioned into two halves. This limitation is common to any stability selection method (e.g., Meinshausen and B$\ddot{\mbox{u}}$hlmann, 2010), because in order to consider stability, due to that there is only one dataset, data re-generating has to be mimicked by some sort of data re-sampling.

Finally, stability selection is becoming popular for cluster analysis (e.g., Fang and Wang, 2012), an example of unsupervised learning. There is no doubt that in any unsupervised learning, the problem of tuning parameter selection is very difficult, because there is no loss function to guide the selection. Maybe stability selection can be used to select tuning parameters in regularized procedures proposed for unsupervised learning.



%

\nolinenumbers

%
%
%

\end{document}